\documentclass[preprint,showpacs,preprintnumbers,amsmath,amssymb]{revtex4}
\usepackage{amsmath}

\usepackage{graphicx}
\usepackage{pdfpages}
\usepackage{dcolumn}
\usepackage{bm}
\usepackage{amsmath} 
\usepackage{amsfonts}
\usepackage{amssymb}
\usepackage{url}
\usepackage{microtype}
\usepackage{graphicx}%

\newcommand{\qed}{\nobreak \ifvmode \relax \else
      \ifdim\lastskip<1.5em \hskip-\lastskip
      \hskip1.5em plus0em minus0.5em \fi \nobreak
      \vrule height0.75em width0.5em depth0.25em\fi}

\begin{document}

\preprint{}
\title{Qubit-Qudit Separability/PPT-Probability Analyses and Lovas-Andai Formula Extensions to Induced Measures}
\author{Paul B. Slater}
 \email{slater@kitp.ucsb.edu}
\affiliation{%
Kavli Institute for Theoretical Physics, University of California, Santa Barbara, CA 93106-4030\\
}
\date{\today}
            
\begin{abstract}
We begin by seeking the qubit-qutrit and rebit-retrit counterparts to the now well-established Hilbert-Schmidt separability probabilities for (the 15-dimensional convex set of) two-qubits of  
$\frac{8}{33} = \frac{2^3}{3 \cdot 11} \approx 0.242424$ and (the 9-dimensional) two-rebits of $\frac{29}{64} =\frac{29}{2^6} \approx 0.453125$.  Based in part on extensive numerical computations, we advance the possibilities of a qubit-qutrit value of 
$\frac{27}{1000} = (\frac{3}{10})^3  =\frac{3^3}{2^3 \cdot 5^3} = 0.027$ and a rebit-retrit one of 
$\frac{860}{6561} =\frac{2^2 \cdot 5 \cdot 43}{3^8} \approx 0.131078$. These four values for $2 \times m$ systems ($m=2,3$) suggest certain numerator/denominator sequences involving powers of $m$, which
we further investigate for $m>3$. Additionally, we find that the Hilbert-Schmidt separability/PPT-probabilities for the two-rebit, rebit-retrit and two-retrit $X$-states all equal $\frac{16}{3 \pi^2} \approx 0.54038$, as well as more generally,
that the probabilities based on induced measures are equal across these three sets of $X$-states. Then, we extend the generalized two-qubit framework introduced by Lovas and Andai from Hilbert-Schmidt measures to  induced ones. For instance, while the Lovas-Andai two-qubit function is $\frac{1}{3} \varepsilon^2 (4 -\varepsilon^2)$, yielding $\frac{8}{33}$, its $k=1$ induced measure counterpart is
$\frac{1}{4} \varepsilon ^2 \left(3-\varepsilon ^2\right)^2$,  yielding $\frac{61}{143} =\frac{61}{11 \cdot 13} \approx 0.426573$, where $\varepsilon$ is a singular-value ratio. We investigate, in these regards, the possibility of extending the previously-obtained ``Lovas-Andai master formula''.
\end{abstract}


\maketitle
\tableofcontents
\section{Introduction}
It is now well-established \cite{slater2017master,lovas2017invariance,milz2014volumes,fei2016numerical,shang2015monte,slater2013concise,slater2012moment,slater2007dyson} that the separability probability with respect to Hilbert-Schmidt  measure
\cite{zyczkowski2003hilbert} of the 15-dimensional convex set of two-qubit states ($N=4$) is $\frac{8}{33}$ and of the 9-dimensional convex set of 
two-rebit states, $\frac{29}{64}$ (with that of the 27-dimensional convex set of two-quater[nionic]bits being $\frac{26}{323}$ [cf. \cite{adler1995quaternionic}], among other still higher-dimensional companion results). (Certainly, one can, however, aspire to a yet greater
``intuitive'' understanding of these results, particularly in some ``geometric/visual'' sense [cf. \cite{szarek2006structure,samuel2018lorentzian,avron2009entanglement,braga2010geometrical,gamel2016entangled,jevtic2014quantum}].) It is of interest to compare/contrast these studies with those other quantum-information-theoretic ones, presented in the recent comprehensive volume of Aubrun and Szarek
\cite{aubrun2017alice}, employing {\it asymptotic geometric analysis}.

By a separability probability, we, of course, mean the ratio of the volume of the separable states to the volume  of all (separable and 
entangled) states with respect to the chosen measure, as proposed, apparently first, by {\.Z}yczkowski, Horodecki, Sanpera and Lewenstein \cite{zyczkowski1998volume} (cf. \cite{petz1996geometries,e20020146,singh2014relative,batle2014geometric}). 

In these regards, let us now present the formulas derived by {\.Z}yczkowski and Sommers
for the total ($N^2-1$-dimensional) Hilbert-Schmidt (HS) volumes of the $N \times N$ (off-diagonal complex-valued) density matrices \cite[eq. (4.5)]{zyczkowski2003hilbert} 
(cf. \cite[eq. (14.38)]{bengtsson2008geometry}),
\begin{equation} \label{ZSComplex}
V_{HS}^{\mathbb{C}}(N)= \frac{\sqrt{N} (2 \pi )^{\frac{1}{2} (N-1) N} \prod_{i=1}^N \Gamma(i)}{\Gamma(N^2) },   
\end{equation}
and their ($\frac{N^2+n-2}{2}$-dimensional) real-valued counterparts \cite[eq. (7.7)]{zyczkowski2003hilbert},
\begin{equation} \label{ZSreal}
V_{HS}^\mathbb{R}(N)=  \frac{\sqrt{N} 2^N (2 \pi )^{\frac{1}{4} (N-1) N} \Gamma[(N+1)/2] \prod_{i=1}^N \Gamma(1+i/2)}{\Gamma(N(N+1)/2) \Gamma(1/2) }.     \end{equation}

Further, Andai alternatively employed Lebesgue measure (yielding results equivalent with the use of the normalization factor, $\sqrt{N} 2^{N(N-1)/2}$ \cite[p. 13648]{andai2006volume} to the Hilbert-Schmidt ones of {\.Z}yczkowski and Sommers), obtaining in the complex case \cite[Thm. 2]{andai2006volume},
\begin{equation} \label{AndaiComplex}
V_{Lebesgue}^\mathbb{C}(N)=\frac{\pi ^{\frac{1}{2} (N-1) N} \Pi_{i=1}^{N-1} i!}{\left(N^2-1\right)!}.
\end{equation}
For  the real case  (we are only immediately interested here in the even dimensions $N=4, 6, 8, 10$), taking $2 l=N$, Andai gave 
\cite[Thm. 1]{andai2006volume},
\begin{equation}  \label{AndaiReal}
V_{Lebesgue}^\mathbb{R}(l)=\frac{2^{-l^2-l} \pi ^{l^2} (2 l)!}{l! \left(2 l^2+l-1\right)!}    \prod_{i=1}^{l-1} (2 i)!.
\end{equation}

Our further arguments could be made here with either of these two sets of formulas. For specificity, we will proceed with the second (Andai/Lebesgue) set (cf. (\ref{MZ1}), (\ref{MZ2})).
\section{Qubit-qutrit analyses} \label{qubitqutritsection}
For the two-qubit ($N=4$) case, we have for the 15-dimensional volume of two-qubit states,
\begin{equation} \label{CN4one}
V_{Lebesgue}^\mathbb{C}(4)    = \frac{\pi ^6}{108972864000} =\frac{\pi^6}{2^9 \cdot 3^5 \cdot 5^3 \cdot 7^2 \cdot 11 \cdot 13}.
\end{equation}
Multiplying this by the associated separability probability $\frac{8}{33}$, we have 
\begin{equation} \label{CN4two}
V_{Sep/Lebesgue}^\mathbb{C}(4)    = \frac{\pi ^6}{449513064000} =\frac{\pi^6}{2^6 \cdot 3^6 \cdot 5^3 \cdot 7^2 \cdot 11^2 \cdot 13}.
\end{equation}
So, we see that the same primes (but to different powers) occur in  the denominators of both volume formulas, while the two numerators remain the same.

Let us now see if we can find analogous behavior in the bipartite ($2 \times 3$) qubit-qutrit ($N=6$) case. On the basis of 2,900,000,000 randomly-generated 
qubit-qutrit density matrices \cite[sec. 4]{al2010random},\cite{zyczkowski2011generating}, we obtained an estimate (with 78,293,301
separable density matrices found) for an associated separability probability of 0.026997690.
(We incorporate the results for one hundred million density matrices reported in  \cite[sec. II]{slater2016invariance}.
Milz and Strunz give a confidence interval of $0.02700 \pm 0.00016$ for this probability \cite[eq. (33)]{milz2014volumes}. A [narrower] $95\%$ confidence interval
based  on our just indicated calculation is $[0.0269918, 0.0270036]$. In the decade-old 2007 paper 
\cite[sec.  10.2]{slater2007dyson}, where the $\frac{8}{33}$ two-qubit conjecture was first formulated, we had advanced a hypothesis of $\frac{32}{1199} =\frac{2^5}{11 \cdot 109} \approx 0.0266889$--subsequently rejected as lying outside the confidence interval reported in \cite[sec.II]{slater2016invariance}. An effort to extend the Lovas-Andai form of analysis
\cite{lovas2017invariance} to the qubit-qutrit and rebit-retrit states has been reported in \cite[App. A]{slater2017master}--but, it now seems, that the separability probabilities reported there were subject to some small, yet not explained,  systematic error.) 

We have for the 35-dimensional volume of qubit-qutrit states,
\begin{equation}
V_{Lebesgue}^\mathbb{C}(6)    =     \frac{\pi ^{15}}{298991549953302804677854494720000000} =
\end{equation}
\begin{displaymath}
\frac{\pi^{15}}{2^{24} \cdot 3^{12} \cdot 5^7 \cdot 7^5 \cdot 11^3 \cdot 13^2 \cdot 17^2 \cdot 19 \cdot 23 \cdot 29 \cdot 31}.
\end{displaymath}
Now, we have found that, for  a separability probability of 
\begin{equation}
\frac{27}{1000} =\frac{3^3}{2^3 \cdot 5^3} =(\frac{3}{10})^3= 0.027,
\end{equation}
we would have the corresponding volume of separable states,
\begin{equation}
V_{Sep/Lebesgue}^\mathbb{C}(6)    =    \frac{\pi ^{15}}{298991549953302804677854494720000000} =
\end{equation}
\begin{displaymath}
\frac{\pi^{15}}{2^{27} \cdot 3^{9} \cdot 5^{10} \cdot 7^5 \cdot 11^3 \cdot 13^2 \cdot 17^2 \cdot 19 \cdot 23 \cdot 29 \cdot 31}.
\end{displaymath}
So, we see that only the powers of 2, 3 and 5 are modified,  closely following the pattern observed ((\ref{CN4one})-(\ref{CN4two})) in the $N=4$ scenario.

A  point to note here is that in the $4 \times 4$ density matrix setting, the positivity of  the determinant of the partial transpose is sufficient for separability to hold \cite{augusiak2008universal}, but not so in the $6 \times 6$ setting. (The partial transpose for an entangled state might  have {\it two} negative eigenvalues \cite{johnston2013non}--but not, we note, in the corresponding $X$-states scenario \cite[App. A]{mendoncca2017maximally}.) This multiple eigenvalue property renders it less directly useful to employ determinantal moments of density matrices and of their partial transposes to reconstruct underlying separability probability distributions, as was importantly done in 
\cite{slater2012moment,slater2013concise}, using ``moment-based density approximants'' \cite{provost2005moment}, based on Legendre polynomials.
\subsection{Induced measures}
Let us now investigate qubit-qutrit scenarios in which the measure employed is {\it not} that induced by tracing over a $K$-dimensional environment, where $K=6$, $k=K-6=0$, as in the Hilbert-Schmidt case, but with $K \neq 6$, $k \neq 0$ \cite{zyczkowski2001induced}. 

For the corresponding induced (lower-dimensional) two-qubit cases, we reported, among others, the formula \cite[eq. (2)]{slater2015formulas} \cite[eq. (4)]{slater2016formulas},
\begin{equation} \label{qubitinduced}
 P^{2-qubits}_k=1-\frac{3 \cdot 4^{k+3} (2 k (k+7)+25) \Gamma{(k+\frac{7}{2}}) \Gamma{(2 k+9)}}{\sqrt{\pi} \Gamma{(3 k+13)}}.
\end{equation}
To obtain the volumes with respect to induced measure, in the two-qubit cases ($N=4)$, we must multiply the complex ($\mathbb{C}$) volume forms
of {\.Z}yczkowski and Sommers (\ref{ZSComplex}) and of Andai (\ref{AndaiComplex}) for $N=4$ by \cite[eq. (3.7)]{zyczkowski2001induced}
\begin{equation}
 \frac{217945728000 (1)_k (2)_k (3)_k \Gamma (k+4)}{\Gamma (4 (k+4))},   
\end{equation}
where the Pochhammer symbol is indicated.
Similarly, for the qubit-qutrit case ($N=6$), we must employ
\begin{equation}
\frac{86109566386551207747222094479360000000 (1)_k (2)_k (3)_k (4)_k (5)_k \Gamma
   (k+6)}{\Gamma (6 (k+6))}.    
\end{equation}
\subsubsection{$k=2$, $K=8$} \label{k2K8}
In the two-qubit case for $k=2$, the formula (\ref{qubitinduced})  gives $\frac{259}{442}=\frac{7 \cdot 37}{2 \cdot 13 \cdot 17} \approx 
0.585973$ (see also (\ref{interpolationk2}) below). Now, of 150,000,000 randomly-generated  qubit-qutrit density matrices with the indicated $k=2$ measure, 23,721,307 had PPT's, yielding an estimated separability probability of 0.15814205.

Among these 23,721,307,  only 171 of them passed the further test for {\it separability from spectrum} presented by Johnston \cite[Thm. 1]{johnston2013separability}. That is, only for 
these 171, did the condition hold that $\lambda_1< \lambda_5 +2 \sqrt{\lambda_4 \lambda_{6}}$, where the $\lambda$'s are the six ordered eigenvalues of 
the density matrices, with $\lambda_1$ being the greatest (cf. \cite[App. A]{slater2017master}).

\subsubsection{$k=1$, $K=7$} \label{k1K7}
In the two-qubit case for $k=1$, the formula (\ref{qubitinduced})  gives $\frac{61}{143}=\frac{61}{11 \cdot 13} \approx 0.4265734$ 
 (see also (\ref{interpolation2}) below). 
Of 171,000,000 randomly-generated qubit-qutrit density matrices for $k=1$, 13,293,906 had PPT's, yielding an estimated separability probability of 0.0777402.
Among these 13,293,906,  only 19 passed the previously-noted (Johnston) test for  separability from spectrum.

\subsubsection{$k=-1$, $K=5$}
In the two-qubit case with $k=-1$, the formula (\ref{qubitinduced}) yields $\frac{1}{14}=\frac{1}{2 \cdot 7}$ \cite[sec. III]{slater2015formulas}.
Now, of 294,000,000 randomly-generated such $6 \times 6$ density matrices, 1,435,605 had PPT's, giving 0.00488301, as a separability probability.
\subsubsection{$k=-2$, $K=4$}
In the two-qubit case with $k=-2$, the associated separability probability must be null, since the ranks of the density matrices are not greater than the dimensions of the reduced systems  \cite{ruskai2009bipartite}. (The value zero is, in fact, yielded by the two-qubit formula (\ref{qubitinduced}) for $k=-2$.) Now, of 330,000,000 randomly-generated  $6 \times 6$ density matrices with $k=-2$, 55,037 had PPT's, giving 0.000166779, as an estimated separability probability. 

At the present stage of our research, we are reluctant to advance specific conjectures for the four immediately preceding
induced-measure qubit-qutrit analyses ($k=2,1,-1,-2$).

\section{Qubit-qudit analyses}
\subsection{$2 \times 4$ case}
In \cite[sec. III.B]{slater2016invariance}, we reported a PPT (positive partial transpose) probability, for the $8  \times 8$ density matrices (viewed as $2 \times 4$ systems) of
0.0012923558, based on 348,500,000 random realizations \cite{al2010random}, 450,386 of them having PPT's.  The associated $95\%$ confidence interval is $[0.0128863, 0.0129609]$. (Milz and Strunz did report  an estimate of 0.0013 \cite[Fig. 5]{milz2014volumes}, but gave no 
associated confidence interval or sample size.)

Let us interestingly note that the numerator of the ($2 \times 2$) two-qubit separability probability $\frac{8}{33}$ is $2^3$, and of the ($2 \times 3$) qubit-qutrit conjecture, $\frac{27}{1000}$ is $3^3$. So, we might speculate that in this $2 \times 4$ setting, the numerator of the PPT-probability would be $4^3 = 64$. Proceeding as in sec.~\ref{qubitqutritsection}, using the Andai Lebesgue volume formula (\ref{AndaiComplex}), with $N=8$,  we did find a 
candidate PPT-probability (but with a numerator of $4^2$) of  $\frac{16}{12375} =\frac{4^2}{3^2 \cdot 5^3 \cdot 11} \approx 0.001292929$.

It would be of interest to try to examine the issue of what proportion of the  $2 \times 4$ PPT-states are, in fact, separable (cf. \cite[sec. IV]{zyczkowski1999volume}), as opposed to bound entangled, using the methodologies recently presented in \cite{qian2018state,Li2018}.
\subsection{$2 \times 5$ case}
We generated 621,000,000 $10 \times 10$ random such density matrices. Of these, 16,205 had a PPT, giving us as
estimated PPT-probability of 0.0000260950. A possible exact value, in line with the noted numerator phenomenon, might be
$\frac{125}{4790016} = \frac{5^3}{2^8 \cdot 3^5 \cdot 5  \cdot 7 \cdot 11} \approx 0.0000260959$.

In a supplementary analysis, for thirty-six  million $10 \times 10$ density matrices, again randomly generated with respect to Hilbert-Schmidt measure, we  found 950 to have PPT's. Among these,  {\it none} passed the further test for {\it separability from spectrum} \cite[Thm. 1]{johnston2013separability}. That is, for 
none, in this 10-dimensional setting, did the condition hold that $\lambda_1< \lambda_9 +2 \sqrt{\lambda_8 \lambda_{10}}$, where the $\lambda$'s are the ten ordered eigenvalues of 
the density matrices, with $\lambda_1$ being the greatest (cf. \cite[App. A]{slater2017master}).

\section{Rebit-retrit analysis} \label{Rebitretritanalysis}
For the two-rebit ($l=2, N=4$) case, we have for the 9-dimensional volume of two-rebit states,
\begin{equation} \label{RN4one}
V_{Lebesgue}^\mathbb{R}(2)    = \frac{\pi ^4}{967680} =\frac{\pi^4}{2^{10} \cdot 3^3 \cdot 5^ \cdot 7}.
\end{equation}
Multiplying this by the established (by Lovas and Andai \cite[Cor. 2]{lovas2017invariance})  separability probability $\frac{29}{64}$, we find
\begin{equation} \label{RN4two}
V_{Sep/Lebesgue}^\mathbb{R}(2)    =\frac{29 \pi ^4}{61931520} =\frac{29 \pi^4}{2^{16} \cdot 3^3 \cdot 5 \cdot 7}.
\end{equation}
So, we see that only the power of 2 is modified, and the exponents of 3, 5 and 7 in the denominators are unchanged.

Let us now see if we can find analogous simple behavior in the rebit-retrit ($l=3, N=6)$ case. On the basis of 3,530,000,000 randomly-generated 
rebit-retrit density matrices \cite[sec. 4]{al2010random}, with respect to Hilbert-Schmidt measure, we obtained an estimate (with 462,704,503
separable density matrices found) for an associated separability probability of 0.1310777629. The associated 
$95\%$ confidence interval is $[0.131067, 0.131089]$.

We have
for the total (20-dimensional) volume of both separable and entangled rebit-retrit states,
\begin{equation}
V_{Lebesgue}^\mathbb{R}(3)    =    \frac{\pi ^9}{1730063650258944000} =\frac{\pi^{9}}{2^{23} \cdot 3^{6} \cdot 5^3 \cdot 7^2 \cdot 11 \cdot 13 \cdot 17 \cdot 19}.
\end{equation}
Then we found that, assuming a very closely fitting separability probability of 
\begin{equation}
\frac{860}{6561} =\frac{2^2 \cdot 5 \cdot 43}{3^8} \approx 0.1310775796,
\end{equation}
we would have
\begin{equation}
V_{Sep/Lebesgue}^\mathbb{R}(3)    = \frac{859 \pi ^9}{11338145138337015398400} =
\end{equation}
\begin{displaymath}
\frac{859 \pi^{9}}{2^{38} \cdot 3^{6} \cdot 5^{2} \cdot 7^2 \cdot 11}.
\end{displaymath}
So, we see that only the powers of 2 and now of 5 in the denominator are again modified.

We note, in the case of $\frac{860}{6561}$, a possible parallism with the conjectured numerators in the qubit-qudit $2 \times m$ cases being powers of $m$, while now in the real cases, the denominators would be.

Let us further observe that the two-rebit counterpart to the two-qubit induced measure formula (\ref{qubitinduced}) 
is \cite[eq. (4)]{slater2015formulas} \cite[eq. (6)]{slater2016formulas},
\begin{equation} \label{rebitinduced}
 P^{2-rebits}_k=1-\frac{4^{k+1} (8 k+15) \Gamma{(k+2)} \Gamma{(2 k+\frac{9}{2}})}{\sqrt{\pi} \Gamma{(3 k+7)}}.
\end{equation}
For $k=0$, we obtain the noted result, $\frac{29}{64}$.
\section{Rebit-redit analyses}
\subsection{$2 \times 4$ case}
We generated  490,000,000 $8 \times 8$ random density matrices with respect to Hilbert-Schmidt ($k=0$) measure. Of these, 12,022,129 had a PPT, giving us as
estimated PPT-probability of 0.02453496. A good fit is provided by $\frac{201}{8192} = \frac{3 \cdot 67}{2^{13}} \approx 0.0245361$. We note, in light of 
our previous analyses, that the denominator $2^{13}$ is obviously also expressible as $4^{6+\frac{1}{2}}$.
\subsection{$2 \times 5$ case}
We generated  620,000,000 $10 \times 10$ random density matrices with respect to Hilbert-Schmidt ($k=0$) measure. Of these, 1,844,813 had a PPT, giving us as
estimated PPT-probability of 0.002975505. A well-fitting candidate PPT-probability is $\frac{29058}{9765625}= \frac{2 \cdot 3 \cdot 29 \cdot 167}{5^{10}} \approx 0.00297554$.
\section{Quaternionic formulas}
Let us also note that in \cite[Thm. 3]{andai2006volume}, Andai presented the quaternionic ($\mathbb{H}$) counterpart,
\begin{equation}
 V_{Lebesgue}^\mathbb{H}(N)=   \frac{\pi ^{N^2-N} (2 N-2)!}{\left(2 N^2-N-1\right)!} \prod_{i=1}^{N-2}(2 i)!,
\end{equation}
of the complex ($\mathbb{C}$) and real ($\mathbb{R}$) volume formulas ((\ref{AndaiComplex}), (\ref{AndaiReal})) given above.
We, then have for the 27-dimensional volume of the two-quaterbit states,
\begin{equation}
 V_{Lebesgue}^\mathbb{H}(4)=   \frac{\pi ^{12}}{315071454005160652800000} =
\end{equation}
\begin{displaymath}
\frac{\pi^{12}}{2^{15} \cdot 3^{10} \cdot 5^5 \cdot 7^3 \cdot 11^2 \cdot 13^2 \cdot 17 \cdot 19 \cdot 23}.
\end{displaymath}
Multiplying by the established separability/PPT-probability (cf. \cite{hildebrand2008semidefinite}) of $\frac{26}{323}$, 
we find
\begin{equation}
 V_{Sep/Lebesgue}^\mathbb{H}(4)=  \frac{\pi ^{12}}{3914156909371803494400000} =
\end{equation}
\begin{displaymath}
\frac{\pi^{12}}{2^{14} \cdot 3^{10} \cdot 5^5 \cdot 7^3 \cdot 11^2 \cdot 13^2 \cdot 17 \cdot 19 \cdot 23}.
\end{displaymath}
We would like to extend our earlier analyses above to the (50-dimensional) ``quaterbit-quatertrit'' setting. But it is 
clearly a challenging problem
to suitably generate sufficient numbers of random $6 \times 6$ density matrices of such a nature (cf. \cite[App. C]{slater2017master}  of C. Dunkl), in order to obtain the needed probability estimates to attempt to closely fit.

We further note that the two-quaterbit counterpart to the two-qubit  and two-rebit induced measure formulas (\ref{qubitinduced}) and (\ref{rebitinduced}), 
is \cite[eq. (3)]{slater2015formulas} \cite[eq. (5)]{slater2016formulas},
\begin{equation} \label{quaterbitinduced}
 P^{2-quaterbits}_k=1-\frac{4^{k+6} (k(k(2 k(k+21)+355)+1452)+2430)\Gamma{(k+\frac{13}{2})} \Gamma{(2 k+15)}}{3 \sqrt{\pi} \Gamma{(3 k+22)}}.
\end{equation}
\section{Two-qutrits}
In \cite[sec. III.A]{slater2016invariance}, we reported an estimated Hilbert-Schmidt PPT-probability  of 0.00010218 for the two-{\it qutrit} states \cite{baumgartner2006state}, based on one hundred million randomly generated density matrices. Following the framework employed above, we have made some limited efforts to find a possible corresponding exact probability. It is by no means clear, however, if one can hope to extend ($2 \times m$) qubit-based results to a fully qutrit setting. (In any case, we did find that the rational value $\frac{323}{3161088}= \frac{17 \cdot 19}{2^{10} \cdot 3^2 \cdot 7^3} \approx 0.00010218$ provides an exceptional fit.) It would be of interest to try to examine the issue of what proportion of the two-qutrit PPT-states are, in fact, separable (cf. \cite{zyczkowski1998volume}) using the methodologies recently presented in \cite{qian2018state,Li2018}.
\section{$X$-states}
We have found that 
the Hilbert-Schmidt separability/PPT-probabilities for both the ($6 \times 6$) rebit-retrit and ($9 \times 9$) two-retrit 
$X$-states to be, somewhat remarkably, equal to that previously reported \cite[p. 3]{dunkl2015separability} for the lower-dimensional ($4 \times 4$) two-rebit $X$-states, that is, $\frac{16}{3 \pi^2} \approx 0.54038$. (The HS two-qubit $X$-states 
separability probability has previously
been shown to equal $\frac{2}{5} = 0.4$ \cite[eq. (22)]{milz2014volumes} \cite[p. 3]{dunkl2015separability}. In \cite[App. B]{slater2017master}, we noted that Dunkl had concluded that the same separability probability did hold for the qubit-qutrit states.)  

We have also found that the equality between two-rebit and rebit-retrit $X$-states separability probabilities continues to hold when the Hilbert-Schmidt measure (the case $k=0$) is 
generalized to the class of induced measures \cite{zyczkowski2001induced,bengtsson2008geometry}. In Fig. 1, we present two equivalent formulas that yield these induced measure two-rebit, rebit-retrit separability probabilities.
\begin{figure}
\includegraphics[page=1,scale=0.9]{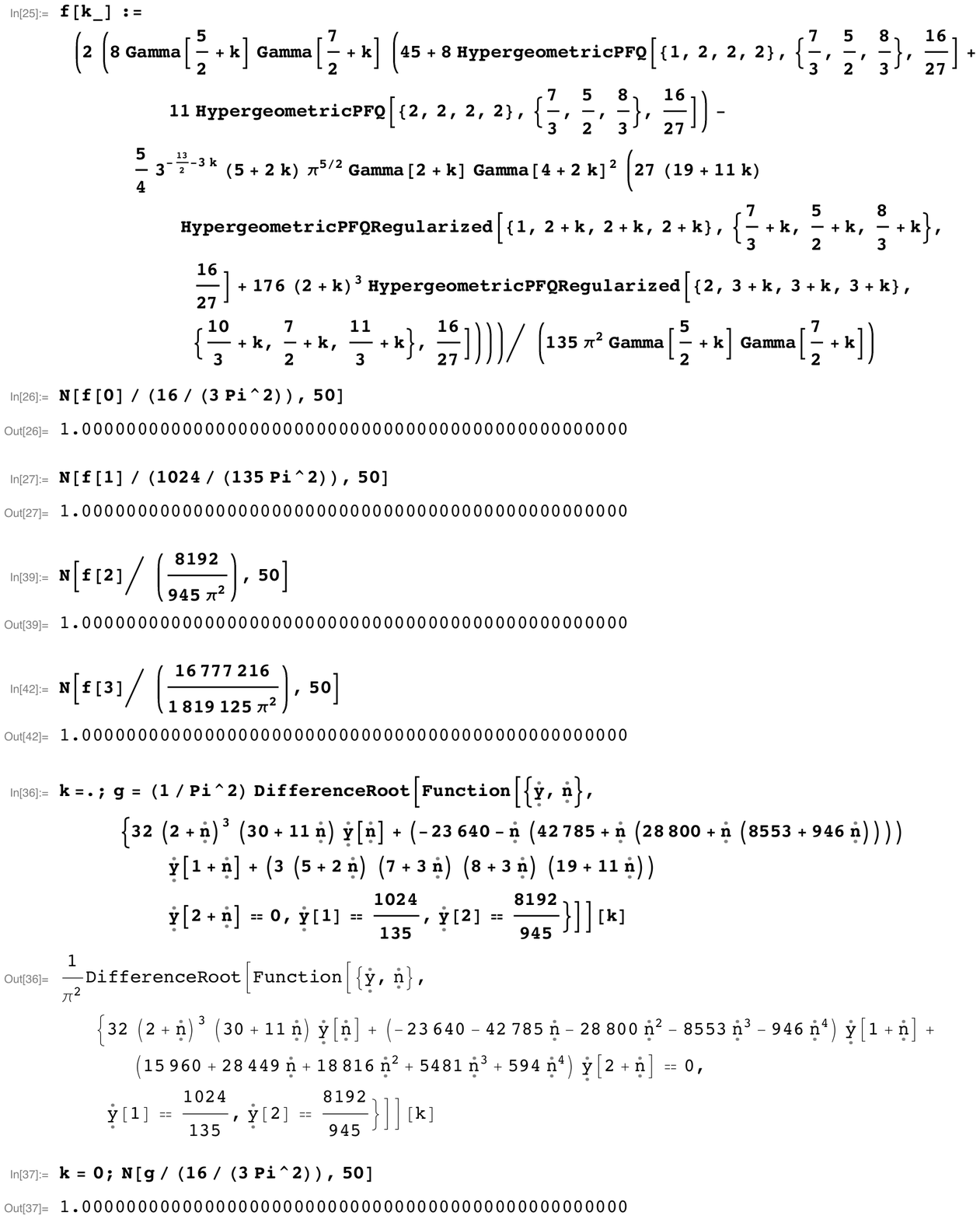}
\label{fig:Induced}
\end{figure}
\begin{figure}
\includegraphics[page=2,scale=0.9]{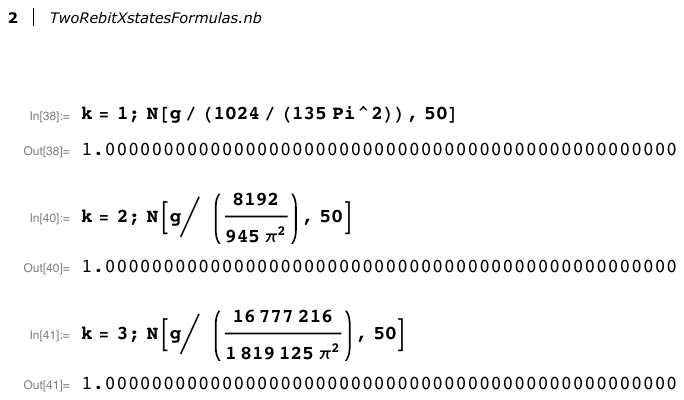}
\caption{Two-rebit and rebit-retrit $X$-states induced separability probability formulas}
\end{figure}
\section{Determinantal equipartition of Hilbert-Schmidt separability probabilities}
In \cite{slater2016formulas}, a formula $Q(k,\alpha)$ was given for that part of the {\it total} induced-measure separability probability, $P(k,\alpha)$, for 
generalized (real [$\alpha=\frac{1}{2}$], complex [$\alpha=1$], quaternionic [$\alpha=2$],...) two-qubit states for which the determinantal inequality $|\rho^{PT}| > |\rho|$ holds. For the 
Hilbert-Schmidt case ($k=0$) the formula yielded $\frac{1}{2}$. (In \cite{szarek2006structure}--making use of Archimedes’ formula for the 
volume of a D-dimensional pyramid of unit height, and of ``pyramid-decomposability''--it was shown that the 
Hilbert-Schmidt separability 
probability of the minimially degenerate states is, likewise, {\it one-half} of that of the nondegenerate states.) Our simulations appear to indicate that this
equal division of separability probabilities continues in the rebit-retrit and qubit-qutrit cases. Based on 96,350,607 separable rebit-retrit cases, the estimated proportion for which  $|\rho^{PT}| > |\rho|$ held was 0.499987, and based on 9,450,652 separable qubit-qutrit cases, the companion estimated proportion  was 0.500033. 

However, in the non-Hilbert-Schmidt analysis in sec.~\ref{k1K7}, pertaining to the qubit-qutrit states with induced measure parameters $k=1$, $K=7$, we
found that the determinantal inequality $|\rho^{PT}| > |\rho|$ held in only $31.17\%$ of the cases. Further, in sec.~\ref{k2K8}, pertaining to the qubit-qutrit states with induced measure parameters $k=2$, $K=8$, the corresponding percentage was $22.63\%$.
\section{Lovas-Andai-type formulas for measures other than Hilbert-Schmidt}
It is of clear interest to extend the forms of analysis above to measures of interest other than the Hilbert-Schmidt
(flat/Euclidean/Frobenius) one, in particular perhaps, the Bures (minimal monotone) one (cf. \cite{slater2000exact}). In these regards, in \cite[sec. VII.C]{slater2017master}, we recently reported, building upon analyses of Lovas and Andai \cite[sec. 4]{lovas2017invariance}, a two-qubit separability probability equal to $1 -\frac{256}{27 \pi^2} =1- \frac{2^8}{3^3 \pi^2} \approx 0.0393251$. This was based on another 
(of the infinite family of) operator monotone functions, namely 
$\sqrt{x}$. (Let us note that the complementary ``entanglement probability'' is simply $\frac{256}{27 \pi^2} \approx 0.960675$. There appears to be no intrinsic reason
to prefer one of these two forms of probability to the other [cf. \cite{dunkl2015separability}].  We observe that the upper-limit-of-integration variable  denoted $K_s:=\frac{(s+1)^{s+1}}{s^s}$, equalling $\frac{256}{27} =\frac{4^4}{3^3}$, for $s=3$, is frequently employed in the Penson-{\.Z}yczkowski paper, ``Product of Ginibre matrices: Fuss-Catalan and Raney 
distributions'' \cite[eqs. (2),  (3)]{penson2011product}.)  
\subsection{Operator monotone measures}
\begin{figure}
    \centering
    \includegraphics{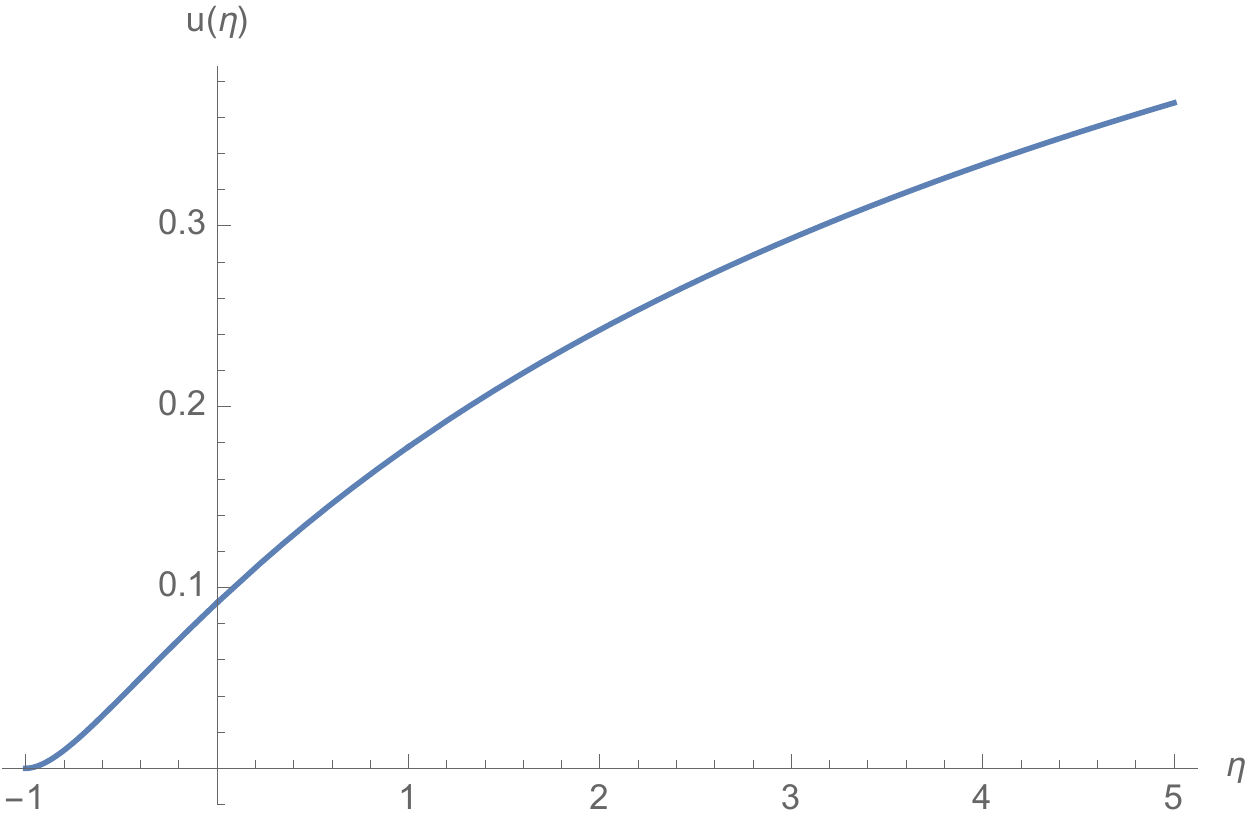}
    \caption{Two-qubit separability probability function $u(\eta)$, given by eq.  (\ref{interpolation}), interpolating between the  ($x \rightarrow \sqrt{x}$) operator monotone result ($\eta=-\frac{1}{2}$) 
    of $1- \frac{256}{27 \pi^2}$ and the Hilbert-Schmidt result ($\eta=2$) of 
    $\frac{8}{33}$.}
    \label{fig:InterpolatedHSmonotone}
\end{figure}

Within the Lovas-Andai framework, employing the previously reported two-qubit ``separability function" $\tilde{\chi}_2(\varepsilon)=\frac{1}{3} \varepsilon^2 (4 -\varepsilon^2)$ 
\cite[eq. (42)]{slater2017master}, we can interpolate between the computation for the noted ($x \rightarrow \sqrt{x}$) operator monotone separability probability of $1- \frac{256}{27 \pi^2}$ ($\eta=-\frac{1}{2}$) 
and the computation for  the Hilbert-Schmidt counterpart of $\frac{8}{33}$ ($\eta=2$).
This is accomplishable using the formula (Fig.~\ref{fig:InterpolatedHSmonotone}),
\begin{equation} \label{interpolation}
u(\eta) =   \frac{\int_{-1}^1  \int_{-1}^x\tilde{\chi}_2 (\sqrt{\frac{1-x}{1+x}}  \sqrt{\frac{1+y}{1-y}})(1-x^2)^\eta (1-y^2)^\eta (x-y)^2 \mbox{d} y \mbox{d} x}{\int_{-1}^1  \int_{-1}^x(1-x^2)^\eta (1-y^2)^\eta (x-y)^2 \mbox{d} y \mbox{d} x}= 
\end{equation}
\begin{displaymath}
-\frac{-3 \eta  (\eta +4) ((\eta -6) \eta -15)+\frac{16^{2 \eta +3} ((\eta -10) \eta -5)
   \Gamma \left(\eta +\frac{3}{2}\right) \Gamma \left(\eta +\frac{5}{2}\right)^3}{\pi ^2
   (2 \eta +3) \Gamma (4 \eta +5)}+60}{3 (\eta -1)^2 \eta ^2},  
\end{displaymath}
where $u(-1)=0$ and $u(1)=\frac{41471}{105}-40 \pi ^2 \approx 0.177729$. 
(It is not now clear if any particularly meaningful measure-theoretic/quantum-information-theoretic interpretation can be given to these interpolated values.)
``We argue that from the separability probability point of view, the main difference between the Hilbert-Schmidt measure and the volume form
generated by the operator monotone function $x \rightarrow \sqrt{x}$ is a special distribution on the unit ball in operator norm of 
$2 \times 2$ matrices, more precisely in the Hilbert-Schmidt case one faces a uniform distribution on the whole unit ball and for 
monotone volume forms one obtains uniform distribution on the surface of the unit ball'' \cite[p. 2]{lovas2017invariance}

Perhaps it is not too unreasonable to anticipate that the Bures two-qubit separability probability 
(associated with the operator montone function $\frac{1+x}{2}$) will also be found to assume a strikingly elegant form. (In 
\cite{slater2002priori}, we had conjectured a value of $\frac{8}{11 \pi^2} \approx .0736881$. But it was later proposed   
in \cite{slater2005silver}, in part motivated by the lower-dimensional {\it exact} results reported in \cite{slater2000exact}, that the value might be $\frac{1680 \sigma_{Ag}}{\pi^8} \approx 0.07334$, where $\sigma_{Ag}= \sqrt{2}-1 \approx 0.414214$ is the ``silver mean''. Both of these studies \cite{slater2002priori,slater2005silver}   were conducted using quasi-Monte Carlo procedures, before the reporting of the  Ginibre-ensemble methodology for generating density matrices, random with respect to the Bures measure \cite{al2010random}.) In 
\cite[sec. VII]{slater2016invariance}, it was noted that ``on the other hand, clear evidence has been provided 
that the apparent $r$-invariance phenomenon revealed by the work of Milz and Strunz,\ldots, does {\it not} continue to hold if one employs, 
rather than Hilbert-Schmidt measure, its Bures (minimal monotone) counterpart''. It would be of interest to examine this issue of $r$-invariance in the context of the induced measures (which, of course, include the Hilbert-Schmidt measure as the special $k=0$ case).
\subsection{Induced measures}
Now, let us raise what appears to be a quite interesting research question. That is, can the  Lovas-Andai framework, which has been successfully applied using both Hilbert-Schmidt and operator monotone function $\sqrt{x}$ measures \cite{lovas2017invariance,slater2017master}, be further adopted to the generalization of Hilbert-Schmidt measure to its induced extensions--through the use of the determinantal powers of density matrices in the derivations? If so, the specific induced separability probabilities reported in 
\cite{slater2015formulas} \cite{slater2016formulas}, including formulas (\ref{qubitinduced}) and (\ref{rebitinduced}) above, as well as 
(\ref{quaterbitinduced}) below, could be presumably further verified. We now investigate this topic.

Let us replace $\tilde{\chi}_2(\varepsilon)=\frac{1}{3} \varepsilon^2 (4 -\varepsilon^2)$  in 
the middle expression in the two-qubit separability probability formula (\ref{interpolation}) for $u(\eta)$ by
\begin{equation}
\tilde{\chi}_{2,1}(\varepsilon)=\frac{1}{4} \varepsilon ^2 \left(3-\varepsilon ^2\right)^2, 
\end{equation}
and set $\eta=3$ (it now being understood, notationally, that $\tilde{\chi}_{2,0}(\varepsilon) \equiv \tilde{\chi}_{2}(\varepsilon)$).

\begin{figure}
    \centering
    \includegraphics{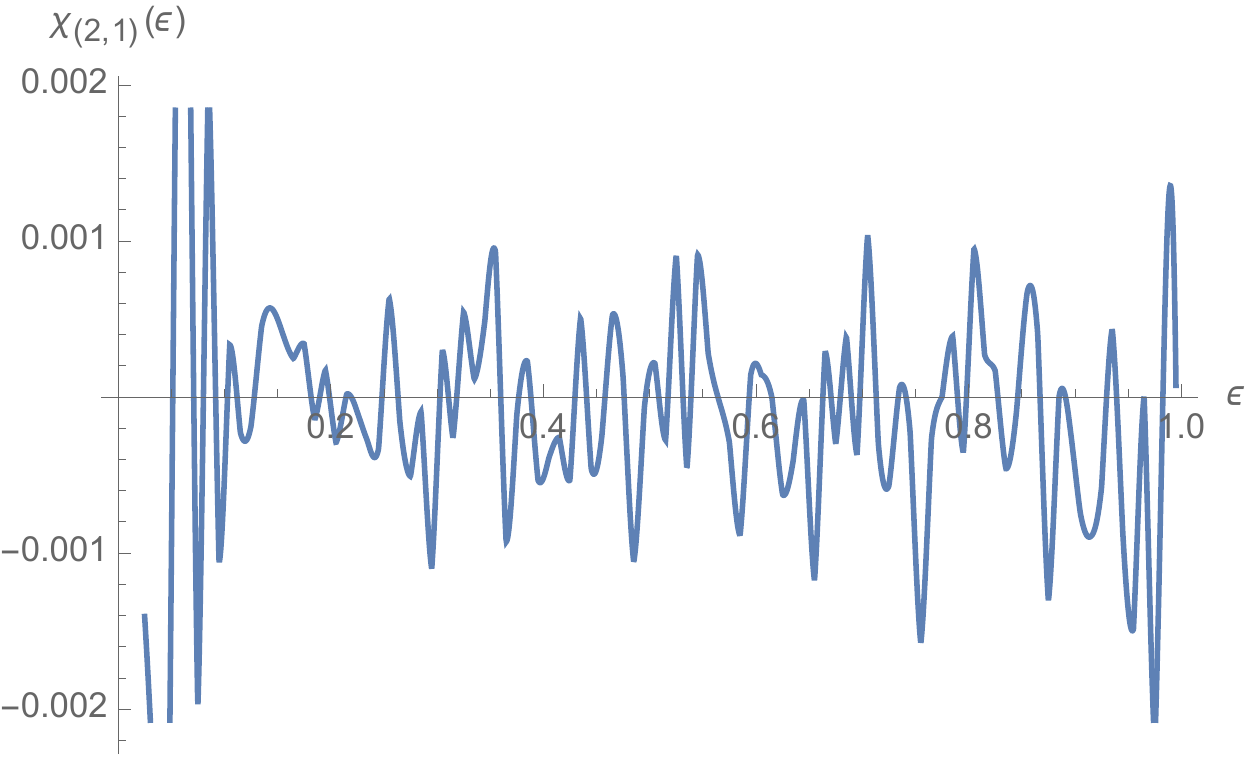}
    \caption{Residuals from a fit of the Lovas-Andai-type formula $\tilde{\chi}_{2,1}(\varepsilon)$ to its estimation based on
sixty million randomly generated ($k=1$) $4 \times 4$ density matrices}
    \label{fig:randominducefitk1}
\end{figure}

Then, this expression does, in fact, evaluate to the two-qubit induced $k=1$ value $\frac{61}{143}$ given by formula (\ref{qubitinduced}).
That is,
\begin{equation} \label{interpolation2}
\frac{\int_{-1}^1  \int_{-1}^x\tilde{\chi}_{2,1} (\sqrt{\frac{1-x}{1+x}}  \sqrt{\frac{1+y}{1-y}})(1-x^2)^3 (1-y^2)^3 (x-y)^2 \mbox{d} y \mbox{d} x}{\int_{-1}^1  \int_{-1}^x(1-x^2)^3 (1-y^2)^3 (x-y)^2 \mbox{d} y \mbox{d} x}=  \frac{61}{143}.
\end{equation}
Fig.~\ref{fig:randominducefitk1} shows the residuals from a (clearly close) fit of  $\tilde{\chi}_{2,1}(\varepsilon)$ to an  estimation of it based on
sixty million appropriately generated $4 \times 4$ density matrices.

Proceeding onward to the $k=2$ case, still in the complex domain ($\mathbb{C}$), we have
\begin{equation} \label{interpolationk2}
\frac{\int_{-1}^1  \int_{-1}^x\tilde{\chi}_{2,2} (\sqrt{\frac{1-x}{1+x}}  \sqrt{\frac{1+y}{1-y}})(1-x^2)^4 (1-y^2)^4 (x-y)^2 \mbox{d} y \mbox{d} x}{\int_{-1}^1  \int_{-1}^x(1-x^2)^4 (1-y^2)^4 (x-y)^2 \mbox{d} y \mbox{d} x}=  \frac{259}{442},
\end{equation}
agreeing with (\ref{qubitinduced}), where, now,
\begin{equation*}
 \tilde{\chi}_{2,2}(\varepsilon)    =\frac{1}{5} \varepsilon ^2 \left(-\varepsilon ^6+8 \varepsilon ^4-18 \varepsilon ^2+16\right).
\end{equation*}
Moving from the complex to quaternionic domain ($\mathbb{H}$), again for $k=1$, we have
\begin{equation} \label{interpolationk3}
\frac{\int_{-1}^1  \int_{-1}^x\tilde{\chi}_{4,1} (\sqrt{\frac{1-x}{1+x}}  \sqrt{\frac{1+y}{1-y}})(1-x^2)^5 (1-y^2)^5 (x-y)^4 \mbox{d} y \mbox{d} x}{\int_{-1}^1  \int_{-1}^x(1-x^2)^5 (1-y^2)^5 (x-y)^4 \mbox{d} y \mbox{d} x}=  \frac{3736}{22287},
\end{equation}
agreeing with (\ref{quaterbitinduced}), where, now, we employ 
\begin{equation}
 \tilde{\chi}_{4,1}(\varepsilon)    = \frac{1}{21} \varepsilon ^4 \left(-9 \varepsilon ^6+55 \varepsilon ^4-125 \varepsilon ^2+100\right).
\end{equation}
(We note that the two-{\it rebit} ($d=1$)  functions $\tilde{\chi}_{1,k}(\varepsilon)$, and more generally $\tilde{\chi}_{d,k}(\varepsilon)$, for odd $d$, appear to be of considerably more complicated non-polynomial form, involving inverse hyperbolic, logarithmic and polylogarithmic functions.)

It now seems clear that to obtain an induced measure-based separability/PPT probability ($\mathcal{P}_{sep/PPT}(d,k)$) in the real ($\mathbb{R}$), complex ($\mathbb{C}$) or quaternionic  
($\mathbb{H}$) domain, we must set the exponent ($d$) of the
$(x-y)$ terms in the numerators and denominators   to 1, 2 or 4, respectively. While to obtain a specific 
$k$-induced measure result, we must take the exponents of the $(1-x^2)$ and $(1-y^2)$ terms to be $d+k$. In other words, we have the general ($(d,k)$-parameterized) 
formula
\begin{equation} \label{interpolationGeneral}
\mathcal{P}_{sep/PPT}(d,k)= \frac{\int_{-1}^1  \int_{-1}^x\tilde{\chi}_{d,k} (\sqrt{\frac{1-x}{1+x}}  \sqrt{\frac{1+y}{1-y}})(1-x^2)^{d+k} (1-y^2)^{d+k} (x-y)^d \mbox{d} y \mbox{d} x}{\int_{-1}^1  \int_{-1}^x(1-x^2)^{d+k} (1-y^2)^{d+k} (x-y)^d \mbox{d} y \mbox{d} x}.
\end{equation}

\begin{figure}
    \centering
    \includegraphics[scale=0.9]{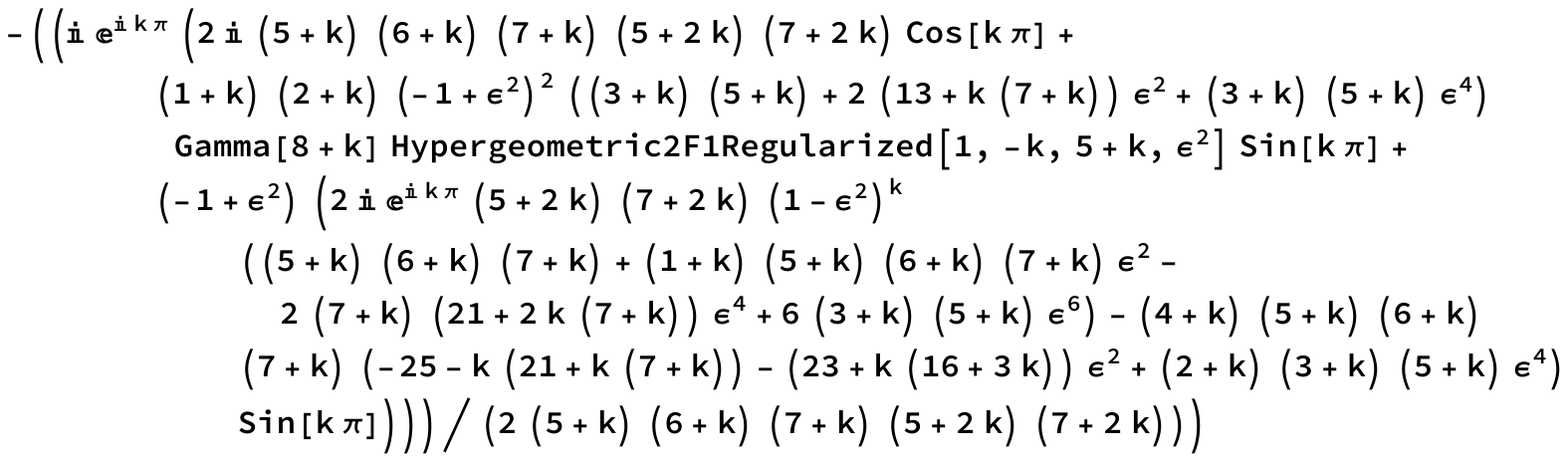}
    \caption{Quaternionic-based ($d=4$) induced-measure formula for $\tilde{\chi}_{4,k}(\varepsilon)$}
    \label{fig:QuaternionicInducedMeasureFormula}
\end{figure}

Now, let us indicate the general manner in which we obtained the three specific indicated new functions $\tilde{\chi}_{2,1}(\varepsilon)$, $\tilde{\chi}_{2,2}(\varepsilon)$ 
and  $\tilde{\chi}_{4,1}(\varepsilon)$ above. 
In this direction, we have for the complex case, $d=2$, the  general induced measure formula
\begin{equation}
 \tilde{\chi}_{2,k}(\varepsilon)   =\frac{\left(-k+\varepsilon ^2-3\right) \left(1-\varepsilon ^2\right)^{k+1}+k+3}{k+3}.
\end{equation}
This gives us for $k=-\frac{5}{2}$,
\begin{equation}
 \tilde{\chi}_{2,-\frac{5}{2}}(\varepsilon)   = 2 \left(\frac{\varepsilon ^2-\frac{1}{2}}{\left(1-\varepsilon
   ^2\right)^{3/2}}+\frac{1}{2}\right),
\end{equation}
which we can interestingly use to replace $\tilde{\chi}_{2}(\varepsilon) $ in (\ref{interpolation}), giving us (again setting $\eta=-\frac{1}{2}$) a result 
now of $\frac{21 \pi -64}{21 \pi } \approx 0.0299127$ to compare (in the induced measure framework) with the previously-given ($x \rightarrow \sqrt{x}$) operator monotone result of 
$1 -\frac{256}{27 \pi^2}  \approx 0.0393251$.
In Fig.~\ref{fig:QuaternionicInducedMeasureFormula}, we present the (quaternionic, $d=4$) formula we have obtained for $\tilde{\chi}_{4,k}(\varepsilon)$. 
For $k=0$, we recover the previously-reported Hilbert-Schmidt formula of $\tilde{\chi}_{4}(\varepsilon)=\frac{1}{35} \varepsilon ^4 \left(15 \varepsilon ^4-64 \varepsilon ^2+84\right)$ 
\cite[sec. VI]{slater2017master}. The corresponding formula for $k=1$ is (\ref{interpolationk3}).

To further elaborate upon the general methodology employed for the above results, we refer to the analyses and notation employed in \cite[sec. VII]{slater2017master}. We must, again, perform the constrained integrations  presented there, but now, additionally,  for induced measure of order $k \neq 0$, we must
multiply both the (numerator and denominator) integrands by the $k$-th power of 
$\left(\left(r_{14}^2-1\right) \left(r_{23}^2-1\right)-r_{24}^2\right) \varepsilon ^2$. This term is the relevant factor in the determinant
of the $4 \times 4$ density matrices (having three pairs of nullified entries) employed in the cited reference. The additional determinantal factors are all positive and not functions of the $r$'s, and would cancel, so can be ignored in the computations.

To be more specific, in these regards, in \cite[sec. VII]{slater2017master}. we employed the set of constraints (imposing--in quantum-information-theoretic terms--the positivity of the density matrix and its partial transpose),
\begin{equation}
r_{23}^2<1\land \left(r_{14}^2-1\right) \left(r_{23}^2-1\right)>r_{24}^2\land r_{23}^2
   \left(\varepsilon ^2 r_{14}^2-1\right)>\varepsilon ^2 \left(\varepsilon ^2
   r_{14}^2+r_{24}^2-1\right).   
\end{equation}
 Then, subject to these constraints, we had to integrate the jacobian (corresponding to the hyperspherical parameterization
of the three off-diagonal non-nullified entries of the density matrix) $\left(r_{14} r_{23} r_{24}\right){}^{d-1}$
over the unit cube $[0,1]^3$. Dividing the result of the integration by
\begin{equation} \label{denominator}
\frac{\pi  4^{-d} \Gamma \left(\frac{d}{2}+1\right)^2}{d^3 \Gamma
   \left(\frac{d+1}{2}\right)^2},
\end{equation}
yielded the desired $\tilde{\chi_d} (\varepsilon )$. (If we were to take $r_{24}=0$, and a jacobian of 
$\left(r_{14} r_{23}\right){}^{d-1}$, we would revert to the X-states setting, and obtain simply 
$\varepsilon^d$ as the corresponding  function.)

This last result (\ref{denominator}) was obtained by integrating the same jacobian 
$\left(r_{14} r_{23} r_{24}\right){}^{d-1}$ over the unit cube,
subject to the constraints (imposing the positivity of the density matrix),
\begin{equation}
r_{23}^2<1\land \left(r_{14}^2-1\right) \left(r_{23}^2-1\right)>r_{24}^2.
\end{equation}
So to reiterate, to move on to the more general induced measure setting (that is, $k \ne 0$), we must multiply both the indicated (numerator and denominator) integrands by the $k$-th power of 
$\left(\left(r_{14}^2-1\right) \left(r_{23}^2-1\right)-r_{24}^2\right) \varepsilon ^2$. The Hilbert-Schmidt ($k=0$) denominator
integration result (\ref{denominator}), then, generalizes to 
\begin{equation}
 \frac{\Gamma \left(\frac{d}{2}\right)^3 \varepsilon ^{2 k} \Gamma (k+1) \Gamma
   \left(\frac{d}{2}+k+1\right)}{8 \Gamma (d+k+1)^2}.   
\end{equation}

\begin{figure}
    \centering
    \includegraphics[scale=0.95]{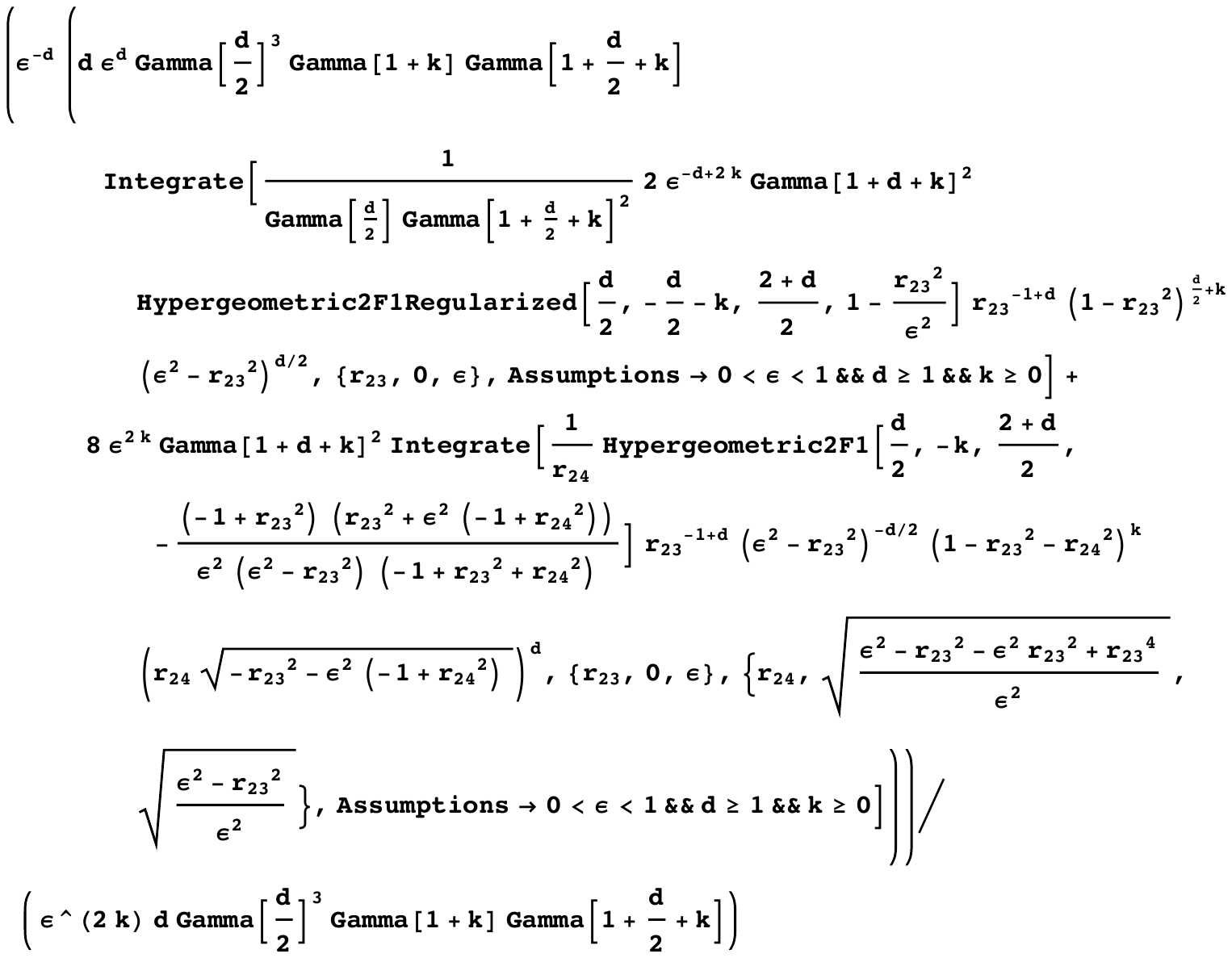}
    \caption{Integration problem for $\tilde{\chi}_{d,k}(\varepsilon)$, which would extend Lovas-Andai (Hilbert-Schmidt)``master formula'' 
    $\tilde{\chi}_{d,0}(\varepsilon) \equiv \tilde{\chi}_{d}(\varepsilon)$, given in (\ref{MasterFormula}), to incorporate induced measures ($k \neq 0$)}
    \label{fig:extendedmasterformula}
\end{figure}

An eventual goal here would be the development of a still more general  Lovas-Andai``master formula'' for $\tilde{\chi}_{d,k}(\varepsilon)$ than has been so far reported for  $\tilde{\chi}_{d}(\varepsilon) \equiv \tilde{\chi}_{d,0}(\varepsilon)$ in \cite[sec.VII.A]{slater2017master}, that is, 
\begin{equation} \label{MasterFormula}
\tilde{\chi}_{d,0}(\varepsilon) \equiv  \tilde{\chi_d}(\varepsilon)= \frac{\varepsilon ^d \Gamma (d+1)^3 \,
   _3\tilde{F}_2\left(-\frac{d}{2},\frac{d}{2},d;\frac{d}{2}+1,\frac{3
   d}{2}+1;\varepsilon ^2\right)}{\Gamma \left(\frac{d}{2}+1\right)^2}.   
\end{equation}
Our efforts in this regard have yielded Fig.~\ref{fig:extendedmasterformula}. The original three-dimensional integration has been reduced
to the sum of a one- and a two-dimensional integration. For $k=0$, (\ref{MasterFormula}) is recovered.
Also, an alternative expression for the anticipated extended master formula is as the sum of
\begin{equation} \label{onehalf}
\frac{(1)_d \varepsilon ^d \Gamma (d+k+1)^2 \,
   _3\tilde{F}_2\left(\frac{d}{2},d,-\frac{d}{2}-k;\frac{d}{2}+1,\frac{3
   d}{2}+k+1;\varepsilon ^2\right)}{d \Gamma \left(\frac{d}{2}\right) \Gamma
   \left(\frac{d}{2}+k+1\right)}    
\end{equation}
(reducing to one-half of (\ref{MasterFormula}) for $k=0$)
and the two-dimensional integral of the product of 
\begin{equation}
  \frac{1}{\Gamma \left(\frac{d}{2}\right)^2 \Gamma (k+1) \Gamma
   \left(\frac{d}{2}+k+1\right)}  
\end{equation}
and
\begin{equation}
Y^{d-1} \left(\frac{1}{r_{14} \epsilon }\right){}^{d+1} \left(1-r_{14}^2 \epsilon
   ^2\right){}^{d/2} \left(\left(1-\frac{1}{r_{14}^2}\right) Y^2-r_{14}^2+1\right){}^k
   \left(r_{14}^2 \epsilon ^2-Y^2\right){}^{d/2}
\end{equation}
and
\begin{equation}
\, _2\tilde{F}_1\left(\frac{d}{2},-k;\frac{d+2}{2};\frac{\left(r_{14}^2 \epsilon
   ^2-1\right) \left(Y^2-r_{14}^2 \epsilon ^2\right)}{\left(r_{14}^2-1\right) \epsilon ^2
   \left(Y^2-r_{14}^2\right)}\right)   .  
\end{equation}
The two-dimensional domain of integration is 
\begin{equation}
r_{14} \in [0,1], \hspace{.25in}  Y \in [\varepsilon   r_{14}, \varepsilon^2   r_{14}]  .
\end{equation}
The result of this integration must also, as (\ref{onehalf}) does, equal one-half of the master formula (\ref{MasterFormula}) result for $k=0$.
Questions pertaining to these last discussed issues have been posted at https://mathematica.stackexchange.com/questions/171351/evaluate-over-a-two-dimensional-domain-the-integral-of-hypergeometric-based-f
and https://math.stackexchange.com/questions/2744828/find-five-parameter-values-for-a-3-tildef-2-function-yielding-five-polynomi      .

\section{Concluding remarks}
Of course, it would be most desirable to  rigorously derive the Hilbert-Schmidt/Lebesgue separability/PPT
probabilities for the 35- and 63-dimensional convex sets of qubit-qutrit and qubit-qudit states, among others,  examined above. But, given that the  Hilbert-Schmidt 
separability probability of $\frac{8}{33}$ for the 15-dimensional convex set of two-qubit  states has itself proved highly formidable to establish \cite{slater2017master,lovas2017invariance,milz2014volumes,fei2016numerical,shang2015monte,slater2013concise,slater2012moment,slater2007dyson}, 
it seems that major advances would be required to achieve such a goal 
in these still higher-dimensional settings (and, thus, confirm or reject the conjectures above).

Implicit in the analytical approach pursued here has been the clearly yet unverified assumption that the separability/PPT-probabilities will continue to be {\it rational-valued} for the higher-dimensional systems, as they have, remarkably, been found to be in the $4 \times 4$ setting.

Our primary goal here  has been to determine if we could use the $N=4$ results \cite{slater2017master,lovas2017invariance,milz2014volumes,fei2016numerical,shang2015monte,slater2013concise,slater2012moment,slater2007dyson} to gain insight into the $N>4$ counterparts, and, more specifically, 
if certain analytical properties continue to hold. We found some encouragement for undertaking such a course from the research reported in 
 \cite{slater2016invariance}. There, evidence was provided that a most interesting common characteristic is 
shared by two-qubit ($N=4$), qubit-qutrit $(N=6$), qubit-qudit ($N=8$, specifically) and two-qutrit ($N=9$) systems. That is, the associated
(HS) separability/PPT probabilities hold constant over the {\it Casimir invariants} \cite{gerdt20116,byrd2003characterization} of both their subsystems
(such as the lengths of the Bloch radii of the reduced qubit subsystems) (cf. \cite[Corollary 2]{lovas2017invariance}). (A Casimir invariant is a  distinguished element of the center of the universal enveloping algebra of a Lie algebra \cite{gerdt20116}.) 

It would be of interest to computationally employ such apparent invariance (formally proved by Lovas and Andai
\cite[Corollary 2]{lovas2017invariance} in the two-rebit $\frac{29}{64}$ case) in strategies to ascertain these various separability/PPT-probabilities. However, we have yet to find
an effective manner of doing so (even after setting the Casimir invariants to zero, leading to 
lower-dimensional settings). (In our paper, ``Two-qubit separability probabilities as joint functions of the Bloch radii of the qubit subsystems'' \cite{slater2016two}, we observed a relative repulsion effect between the Casimir invariants of the two reduced systems of several forms of bipartite states.)

Let us, in these regards, also indicate the interesting paper of Altafini, entitled ``Tensor of coherences parametrization of multiqubit density operators for entanglement characterization'' \cite{altafini2004tensor}. In it, he applies the term ``partial quadratic Casimir invariant'' in relation to reduced density matrices. He notes that a quadratic Casimir invariant can be regarded as the specific form ($q = 2$) of Tsallis entropy. Further, he remarks that ``partial transposition is a linear norm preserving operation: $\mbox{tr}(\rho^2) = \mbox{tr}((\rho^{T_1})^2) =  \mbox{tr}((\rho^{T_2})^2)$. Hence entanglement violating PPT does {\it not modify} the quadratic Casimir invariants of the density and the necessary [separability] conditions $[\mbox{tr}(\rho^2_A) \geq \mbox{tr}(\rho^2), \mbox{tr}(\rho^2_B) \geq \mbox{tr}(\rho^2)] $, are 
{\it insensible} to it'' (emphasis added).

Let us, relatedly, indicate the pair of formulas (cf. (\ref{ZSComplex}), (\ref{AndaiComplex}))
\begin{equation} \label{MZ1}
  V_{HS}^{2 \times m}(r)=   V_{HS}^{2 \times m}(0) (1-r^2)^{2 (m^2-1)}
\end{equation}
and
\begin{equation} \label{MZ2}
 V_{HS}^{2 \times m}(0)   = \sqrt{m} \cdot 2^{6 m^2 -m -\frac{23}{2}} \cdot \pi^{2 m^2-m -\frac{3}{2}} \cdot 
 \frac{\Pi_{k=1}^{2 m} \Gamma{(k)} \cdot \Gamma{(\frac{1}{2}+2 m^2)}}{\Gamma{(4 m^2)} \cdot \Gamma{(-1+2 m^2)}}
\end{equation}
that Milz and Strunz conjectured for the Hilbert-Schmidt volume of the $2\times m$ qubit-qudit states \cite[eqs. (27), (28)]{milz2014volumes}, as a function of the Bloch radius ($r$) of the qubit subsystem. (These appear to have been confirmed for the two-qubit 
[$m=2$] case by the analyses of Lovas and Andai 
\cite[Cor. 1]{lovas2017invariance}.)

We can, of course, as future research, continue our simulations of random density matrices, hoping to obtain further accuracy in our
various separability/PPT-probability estimates. One relevant issue of interest would then be the trade-off between the use of  increased precision in the random normal variates employed (we have so far used the Mathematica default option), and the presumed consequence, then, of decreased number of  variates to be generated.

\bibliography{main}

\end{document}